\documentclass[letterpaper,12pt]{article}

\usepackage{titlesec}
\titleformat*{\section}{\bfseries} 
\titleformat*{\subsection}{\bfseries} 

\usepackage[top=1in, bottom=1in, left=1in, right=1in]{geometry}
\usepackage{amssymb}
\usepackage{graphicx}
\usepackage{hyperref}
\usepackage{natbib}
\usepackage{aas_macros}

\newcommand{\someskip}{\bigskip}

\begin{document}

\noindent {\bf Astro2020 Science White Paper}

\someskip

\noindent {\bf Exploring Active Supermassive Black Holes at 100 Micro-arcsecond Resolution}

\someskip

\noindent {\bf Thematic Areas:} Galaxy Evolution

\someskip

\noindent {\bf Principal Author:}

\noindent Name: Makoto Kishimoto

\noindent Institution: Kyoto Sangyo University

\noindent Email: mak@cc.kyoto-su.ac.jp

\noindent Phone: +81-75-705-3039

\someskip

\noindent {\bf Co-authors:} (names and institutions)

\noindent Theo ten Brummelaar (The CHARA Array of Georgia State University)

\noindent Douglas Gies (Georgina State University)

\noindent Robert Antonucci (University of California, Santa Barbara)

\noindent Sebastian H\"onig (University of Southampton)

\noindent Martin Elvis (Harvard-Smithsonian Center for Astrophysics)

\noindent John Monnier (University of Michigan)

\noindent Stephen Ridgway (NOAO)

\noindent Michelle Creech-Eakman (New Mexico Tech)

\someskip

\noindent {\bf Abstract:} Super-high spatial resolution observations in the infrared are now enabling major advances in our understanding of supermassive black hole systems at the centers of galaxies. Infrared interferometry, reaching resolutions of milliarcseconds to sub-milliarcseconds, is drastically changing our view of the central structure from a static to a very dynamic one by spatially resolving to the pc-scale. We are also starting to measure the dynamical structure of fast moving gas clouds around active supermassive black holes at a scale of less than a light year. With further improvements on resolution and sensitivity, we will be able to directly image the exact site of the black hole's feedback to its host galaxy, and quantify the effect of such interaction processes. Near-future high angular resolution studies will definitely advance our mass determinations for these black holes, and we might even witness the existence of binary black hole systems at the center of galaxies.

\newpage 

\section{An emerging new picture for AGN structure}
\label{sect-torus}

Over the last few years, the study of active galactic nuclei (AGNs) has been going through a transformation. In the standard picture, which has been around for more than 30 years (\citealt{Antonucci85}), an obscuring equatorial `torus' is invoked and believed to surround the accreting supermassive black hole at the center. The physical origin of this torus has not been clear, but it is assumed to be more or less static. Its existence unifies the two major AGN categories: those with a face-on, polar, direct view of the nucleus, called Type 1, and those with an edge-on, equatorial view, with the nucleus hidden, called Type 2. 

\begin{figure}[!ht]
\includegraphics[width=1.0\textwidth]{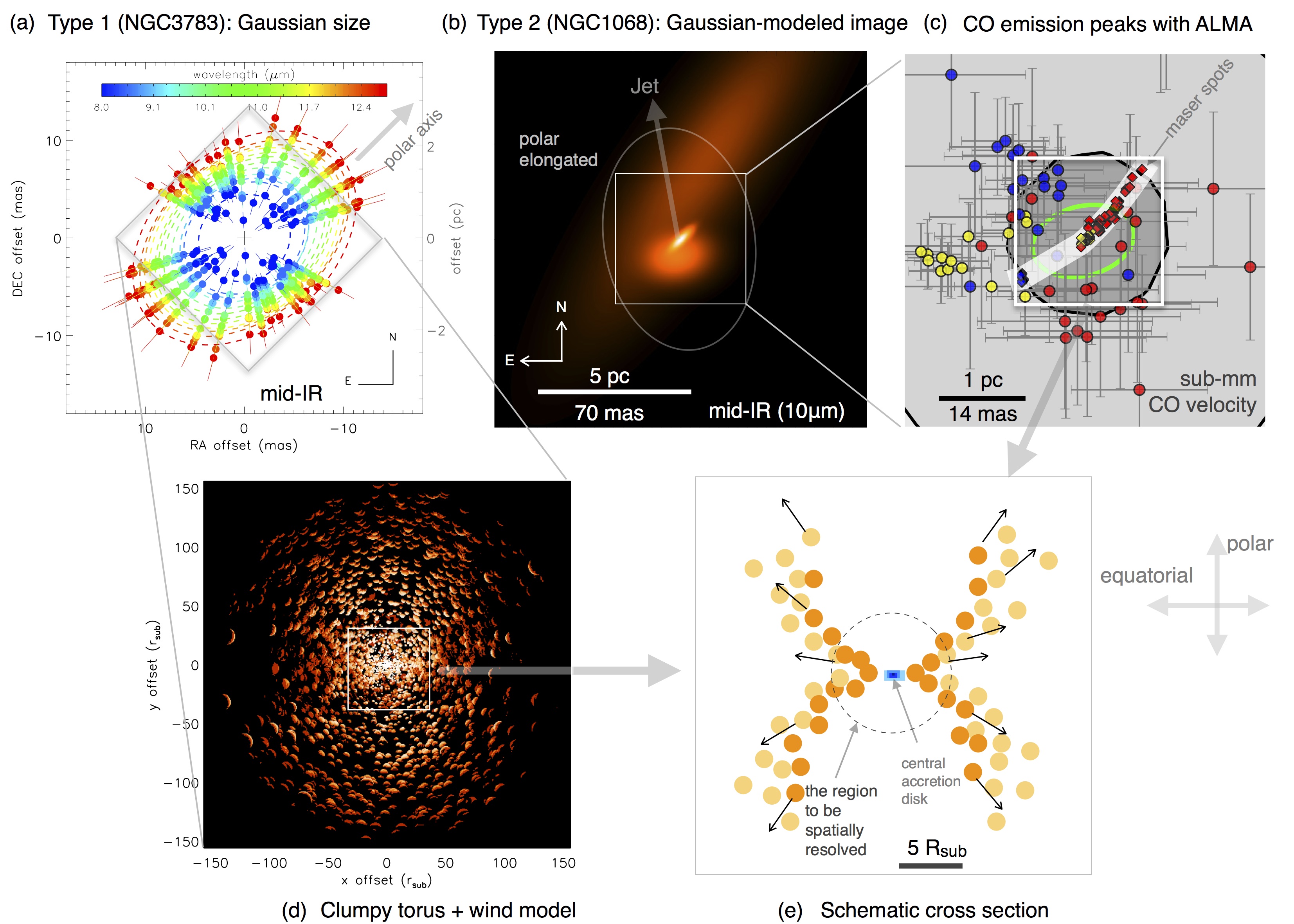}
\caption{\small 
{\bf (a)} Mid-IR Gaussian size observed for the southern brightest Type 1 AGN NGC3783 (from \citealt{Hoenig13}). The polar axis direction inferred from optical polarization measurements is indicated.
{\bf (b)} Three-component Gaussian model image of NGC1068 based on mid-IR interferometry data (from \citealt{LopezGonzaga14}), with the linear jet (i.e.,~polar) direction indicated. 
{\bf (c)} CO emission peaks observed in the sub-mm, indicated by circles and color-coded with velocities ($<$$-$70 km/s in blue; $>$+70 km/s in red; from \citealt{Gallimore16}). 
{\bf (d)} Clumpy torus + wind model for a Type 1 inclination of 25$^{\circ}$ (from \citealt{Hoenig17}).
{\bf (e)} Schematic picture of an obscuring and outflowing torus.
}
\label{windy-torus}
\end{figure}

However, recent mid-IR interferometry has shown that a major part of the mid-IR emission, believed to be from the outer warm ($\sim$300K) part of this putative dusty torus, has a {\em polar-elongated} morphology, rather than the expected equatorially elongated structure (e.g., \citealt{Hoenig12,Hoenig13,LopezGonzaga14}; Fig.\ref{windy-torus}a,b). In addition, this polar-elongated dusty gas is in fact considered to be UV-optically-thick, since the measured IR emissivity is 0.2-0.3 and consistent with directly illuminated UV-optically-thick gas (e.g., Fig.3 in \citealt{Hoenig10model}). Furthermore, at the same spatial scale, ALMA is finding a polar outflow (\citealt{GarciaBurillo16,Gallimore16}; Fig.\ref{windy-torus}c), likely to be an inward extension of the 10-100~pc scale bipolar outflow directly resolved by HST (e.g., \citealt{Cecil02}).

Therefore it is quite likely that there is a nuclear, polar outflow which is UV-optically-thick, i.e., participating in obscuring the nucleus. In fact, the interferometric data showing the polar elongation of Type 1, and the whole infrared spectral energy distribution (SED), can be simultaneously modeled by a clumpy torus + wind model (\citealt{Hoenig17}; Fig.1d). Thus, a new observational scenario emerging here is that the torus is actually an {\it obscuring outflow}, with the polar part hollow, in order to have a Type-1 line-of-sight unobscured. This structure is probably driven by radiation pressure on dust grains from the anisotropic, polar-strong UV radiation from the central accretion disk, as illustrated in Fig.1e. This picture is quite consistent with the results of hydrodynamical simulations (\citealt{Wada12}).

This dusty wind is likely giving strong feedback to the host galaxy, which could be regulating the strong correlation between the bulge and black hole mass in galaxies (e.g., \citealt{Ferrarese00}). This would mean that the `torus' - a notion which has been around for so many years - might actually be the exact location of the black hole's feedback to galaxy evolution. We might be able to directly quantify this feedback process by spatially resolving the region itself. 

The key for this big advance is the super high spatial resolution -- the torus has been extensively investigated with SED modeling efforts based on a `traditional' equatorial torus picture, but the high angular resolution enabled by IR interferometry is now radically changing the view of this torus structure. The current and coming instrumentation providing high spatial resolution in the thermal infrared emission should thus have considerable discovery potential.

\section{AGN broad-line region}

Further down in the spatial scale inside the `torus', fast moving clouds are believed to reside in a relatively low density region around the central supermassive black hole, emitting very broad emission lines - the so-called broad-line region (BLR). The broadening of the lines has been believed mainly due to the fast orbital motion of the clouds. Black hole masses in AGNs have been inferred mostly from this supposed rotational speed of the clouds indicated by the width of the lines, thus the region has been playing a major role in Astronomy. However, the assumed `ordered' rotation has never been directly shown until quite recently. Here, a big advance has been achieved by GRAVITY, one of the second-generation instruments at VLTI.

\begin{figure}[!t]
\centering
\includegraphics[width=6in]{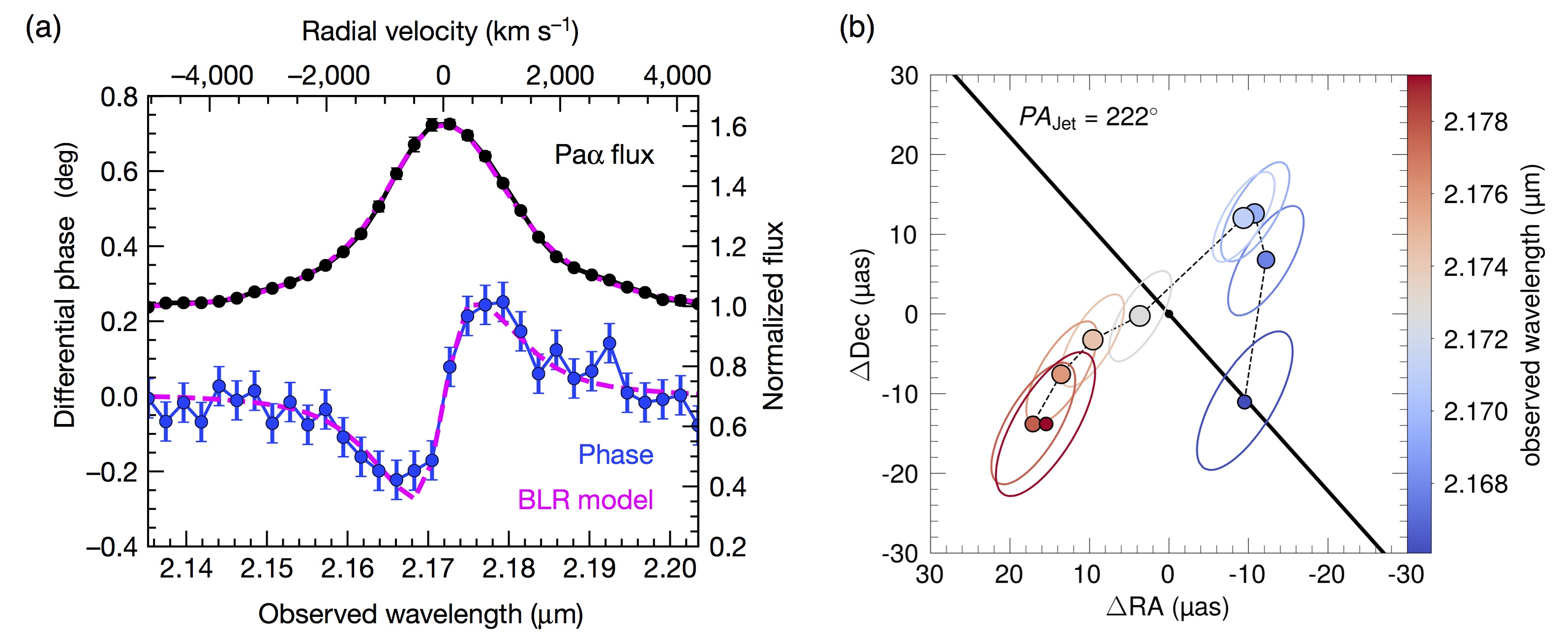}
\caption{\small 
{\bf (a)} Differential phase spectra of 3C273, showing a systematic phase shift across the P$\alpha$ line.
{\bf (b)} Photo-center at each wavelength bin over P$\alpha$ line wavelengths with ellipticals representing uncertainties.
Both figures are from \cite{GravitySturm18}.
}
\label{gravity}
\end{figure}

GRAVITY interferometrically combines the beams from the four 8.2~m telescopes at the VLT site, and it can now produce images at a few mas (milli-arcsecond) resolution. This is already fantastic, but this resolution in fact is not good enough for spatially resolving the BLR. Fortunately, the instrument can accurately measure the variation of interferometric phase with wavelength, meaning that it can measure the relative position of the image photo-center within emission lines. Using this capability, the instrument has now captured the BLR of the brightest quasar on the sky, 3C273 at redshift 0.156 -- measuring photo-center displacements across the broad P$\alpha$ line at $\sim$10 {\it micro}-arcsecond (Fig.\ref{gravity}) precision. From the red side to the blue side of the line, the photo-center gradually moves perpendicularly to the system polar axis (known from the jet), indicating an ordered rotation of the BLR (perhaps except for the bluest point; see sect.\ref{sect-line} below). This really shows that IR interferometry has reached a state of maturity where we can now explore the innermost regions of AGNs in exquisite detail.

\begin{figure}[t]
\includegraphics[width=\textwidth]{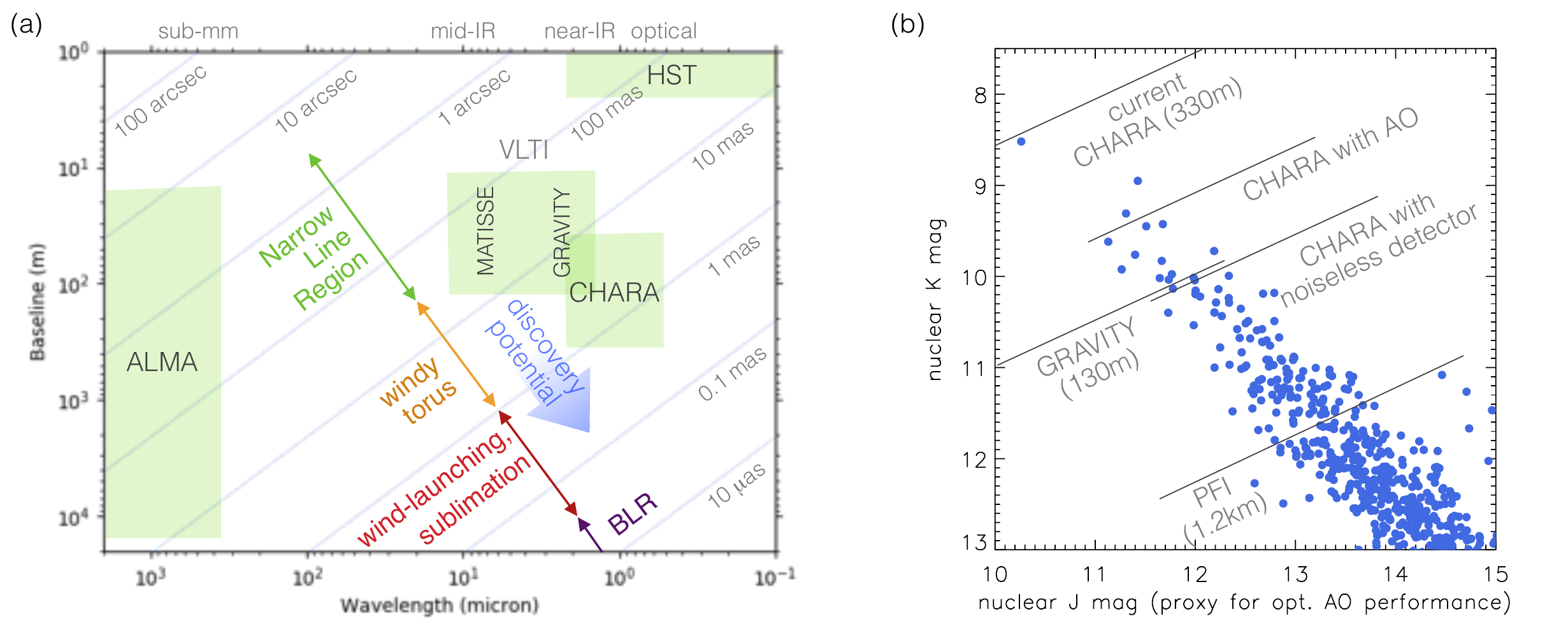}
\caption{\small 
{\bf (a)} Spatial resolutions of current facilities are shown on the plane of observing wavelength $\lambda$ and baseline $b$, where the dotted lines indicate the limit given by $\lambda/b$. Approximate scales of various regions for nearby AGNs are also shown.
{\bf (b)} AGN targets on the plane of nuclear K mag vs nuclear J mag. The latter approximately represents the optical AO performance. Current and future  interferometric observations' limits are indicated.
}
\end{figure}

\section{Quest for sub-milliarcsec resolution imaging in the infrared}

\subsection{Morphological studies using continuum}

In the new emerging picture of AGNs described in section~\ref{sect-torus}, the outflow is understood to be driven by the radiation pressure on dust, which has been considered as a plausible candidate for the acceleration source (e.g., \citealt{Fabian08}). In this case, the wind launching region is the innermost dusty region, or the dust sublimation region, and this site must be the origin of the AGN feedback on the ambient gas in the host galaxy. We argue here that we will be able to spatially resolve and scrutinize this critical region, and possibly quantify the effect of such interaction processes, if we push our resolving power and sensitivity further.

Since the dust sublimation temperature is expected to be $\sim$1500 K, the continuum emission of the region must be brightest in the near-IR ($\sim$2 $\mu$m). Over the last several years, intensive efforts have been made to spatially resolve this region with interferometers in the near-IR. With the Keck interferometer and the first generation instrument AMBER at VLTI having $\sim$100 m baselines, the region has been {\em marginally} resolved -- the slight visibility drops show that the overall size of the region is $\sim$1 mas or less for the brightest AGNs (\citealt{Kishimoto11, Kishimoto13KI, Weigelt12}). This is depicted in Fig.3a, which shows the approximate size scale of each layer of the structure and compares various instruments on the plane of wavelength and baseline.

The second-generation instrument GRAVITY at VLTI is gradually reaching the stage to resolve at least the outer part of the region, with a real imaging capability at a resolution of a few mas. The CHARA interferometer, which has a factor of $\sim$3 longer baselines (330~m), but with 1-m telescopes, has already seen a first fringe for an AGN, and the installation of full adaptive optics is now in progress. If it goes through further, already proposed, sensitivity enhancements, we will be able to image the wind-launching region of a few AGNs. The study will be complemented by observations at slightly longer wavelengths in the mid-IR probing slightly lower temperature dust with the newly commissioned second-generation mid-IR interferometer MATISSE at VLTI (Fig.3a). Then, with longer-baseline interferometers, reaching resolutions of 100 micro-arcseconds and beyond, such as the Planet Formation Imager (PFI) planned both for mid-IR and near-IR with higher sensitivity, we can study a substantial sample. We will know whether the dust continuum emission shows a polar elongation even at this innermost scale, gain knowledge of how the wind is launched, and fully characterize this critical scale, approaching the origin of the AGN torus and feedback.

\subsection{Kinematical studies using emission lines}
\label{sect-line}

Together with the morphological continuum study, we should also identify the outflow kinematically, using emission lines from the ionized gas associated with the accelerated dust grains. Any velocity gradient detection spatially coincident with a polar elongation would greatly advance our knowledge. Such observations will also help quantify the feedback effect. 

In fact, with the photo-center shifts from the GRAVITY differential phase data, in addition to the ordered rotation, we might actually be seeing a hint of high-velocity outflow along the jet (see the bluest data point in Fig.2b), though with a large uncertainty. At least, imaging of emission-line kinematics at this super high spatial resolution is becoming realistic. Note that this is the data for a broad permitted line, i.e., for the BLR --- the torus outflow might thus be fundamentally and physically related to the BLR. Spatially resolving these scales will definitely advance the way we measure the black hole masses. We would need very long baselines to reach these scales, and high sensitivities to receive enough number of photons to facilitate good spectral resolution for kinematics (Fig.3a,b), but we do believe that the discovery potential lies here at high angular resolutions.

The morphology and kinematics of the bright, illuminated side of the torus would be nicely complemented by molecular line studies of the ``dark" side with ALMA, which comes quite close in spatial scales at its longest baseline (Fig.3a). The torus should also be supplying material to the central black hole, and the route is perhaps at the mid-plane. In fact, an equatorial rotation component is found with high-density tracers of molecular lines (\citealt{Imanishi18}) in addition to the lower-density polar outflow (\citealt{GarciaBurillo16,Gallimore16}).

Current sensitivity of near-IR interferometry is shown in Fig.3b on the plane of nuclear K-band and J-band magnitudes, the latter representing coarsely the quality of the AO performance which is a must in IR interferometry. The distribution of AGNs on this plane demonstrates that we are reaching the ``tipping point": the number of potential targets will soon explode with the future sensitivity improvements --- and note that we will have {\it images} --- we are at the dawn of the field.

\section{Binary supermassive black holes}

There is also a substantial, but very different, aspect to this new exploration of the galactic nuclei at a very high angular resolution. A supermassive black hole is now believed to reside in essentially every galaxy. Naturally, after collision and merger of two such galaxies, a binary black hole would form at the center. Theoretical studies with numerical simulations have suggested that the binary separation quickly decreases by ejection of surrounding stars. However, the shrinking orbit could stall at 0.1 pc scale owing to the absence of a sizable population of stars in the small region -- this is known as the ``final parsec problem" (e.g., \citealt{Merritt05}). Indeed, this inference seems consistent with recent Pulsar Timing Array constraints: low-frequency gravitational waves, expected from supermassive black hole mergers, are not yet detected (e.g., \citealt{Shannon15}). 

This spatial scale of $\sim$0.1 pc is exactly the scale which has not been explored in thermal emission. A radio wavelength search for binary structure relies on the presence of two sets of jets, which might not necessarily always be the case. We do not have enough spatially resolved constraints in thermal emission simply because it has not been possible to achieve the required resolving power. We do not really know if there is a binary at the center in the active phase of the black hole accretion. Long-baseline interferometry in the infrared will enable us for the first time to explore this major question of whether there are massive binaries at the center.

\bibliographystyle{aasjournal}

\end{document}